\documentclass[pre]{revtex4}
\usepackage{amsmath,amssymb,latexsym,epsfig,graphics,epsf}
\usepackage{graphics}
\usepackage{float}
\newcommand{\br}{\bold r}

\newcommand{\trh}{\tilde{\rho}}
\newcommand{\tT}{\tilde{T}}
\newcommand{\tV}{\tilde{V}}

\newcommand{\bR}{\bold R}
\newcommand{\tO}{\tilde\Omega}
\newcommand{\tbR}{\tilde{\bold R}}
\newcommand{\tU}{\tilde{\Phi}}

\newcommand{\be}{\begin{equation}}
\newcommand{\ee}{\end{equation}}

\begin{document}
\title{Isomorphs, hidden scale invariance, and quasiuniversality}
\author{Jeppe C. Dyre}\email{dyre@ruc.dk}
\affiliation{DNRF Center "Glass and Time", IMFUFA, Dept. of Sciences, Roskilde University, P. O. Box 260, DK-4000 Roskilde, Denmark}

\date{\today}

\begin{abstract}
This paper first establishes an approximate scaling property of the potential-energy function of a classical liquid with good isomorphs (a Roskilde-simple liquid). This ``pseudohomogeneous'' property makes explicit that -- and in which sense -- such a system has a hidden scale invariance. The second part gives a potential-energy formulation of the quasiuniversality of monatomic Roskilde-simple liquids, which was recently rationalized in terms of the existence of a quasiuniversal single-parameter family of reduced-coordinate constant-potential-energy hypersurfaces [J. C. Dyre, Phys. Rev. E {\bf 87}, 022106 (2013)]. The new formulation involves a quasiuniversal reduced-coordinate potential-energy function. A few consequences of this are discussed.
\end{abstract}
\maketitle

\section{Introduction}\label{intro}

Traditionally a liquid is termed simple if it consists of point-like particles interacting via radially symmetric pair potentials \cite{fis64,ric65,tem68,ail80,rowlinson,gubbins,chandler,barrat,deb05,dou07,kir07,bag10,han13}; a prime example is the Lennard-Jones system. Computer simulations during the last 20 years have revealed, however, that a number of such systems, e.g., the Gaussian core model, the Lennard-Jones Gaussian model, and the Jagla model have quite complex properties (see Ref. \onlinecite{dyr13} and its references). On the other hand, many {\it molecular} models exhibit simple behavior in simulations, and experiments on van der Waals bonded molecular liquids show that these are generally regular with no anomalous behavior. These facts led us recently with Ingebrigtsen and Schr{\o}der to suggest defining a simple liquid as one with strong correlations between the equilibrium virial and potential-energy fluctuations in the canonical ($NVT$) ensemble \cite{ing12}, a condition that is obeyed by Lennard-Jones type liquids and many other systems \cite{I,ing12,ing12b}. In this definition the degree of simplicity generally depends on the thermodynamic state point in question; in fact all realistic systems lose simplicity approaching the critical point and gas states. We originally used the term ``strongly correlating liquids'' for these systems, but that often led to confusion with strongly correlated quantum systems. Terming the class ``simple liquids'' has been criticized for the risk of mistaking it for the traditional simplicity concept, so we now refer to the class in question as ``Roskilde-simple liquids'' \cite{sep13,ing13,dyr13a}.

Appendix A of Ref. \onlinecite{IV} established a constitutive theorem giving three characterizations of Roskilde-simple liquids. It states that a system has strong equilibrium $NVT$ virial potential-energy correlations if and only if it has good isomorphs, which are curves in the thermodynamic phase diagram along which structure and dynamics are invariant in reduced units \cite{IV}. Moreover, this happens if and only if the system has curves in the phase diagram (the isomorphs) along which the reduced-coordinate constant-potential-energy hypersurfaces are invariant. The latter property led to the introduction of $NVU$ dynamics \cite{NVU_I,NVU_II,NVU_III} -- geodesic motion on the constant-potential-energy hypersurface. This is a novel molecular dynamics that in the thermodynamic limit for virtually all structural and dynamic properties gives results identical to those of conventional Newtonian dynamics. 

Only systems for which the potential energy is an Euler homogeneous function have 100\% virial potential-energy correlation and perfect isomorphs, and for all realistic Roskilde-simple liquids the isomorph concept is only approximate. Extensive computer simulations have shown \cite{IV,sep13,ing13,dyr13a,ing12,V,gna10,ing12b,ing12a,vel12,boh12,ped10}, however, that the predicted isomorph invariants apply to a good approximation for a wide variety of systems. A few experimental predictions of the isomorph theory have been confirmed, as well \cite{gun11,boh12,roe13}. 

This paper first discusses the case of general Roskilde-simple liquids, including mixtures \cite{IV,vel12,ped10} and molecular systems \cite{ing12b}. A simple approximate scaling property of the potential-energy function is established from which the entire isomorph theory \cite {IV} is easily derived. The second part of the paper deals with the quasiuniversality of monatomic Roskilde-simple liquids. Recently we showed \cite{dyr13} that the several intriguing quasiuniversalities reported during the last 50 years may all be regarded as consequences of one single quasiuniversality: the family of reduced-coordinate constant-potential-energy hypersurfaces is quasiuniversal. It is shown below that this implies that all Roskilde-simple liquids have basically the same potential-energy function. More accurately, the potential energy of one such liquid, $U_1$, is to a good approximation a linear function of the potential energy of any other, $U_1\cong\alpha U_2+\beta$, in which the two constants are functions of density. From this one may derive all the quasiuniversalities. 

The paper starts by giving preliminaries and summarizing briefly the theory of isomorphs (Sec. \ref{backgr}). Section \ref{gencase} demonstrates the pseudohomogeneous nature of the potential-energy function of a Roskilde-simple liquid by establishing an approximate scaling expression. Based on this Sec. \ref{hsi} discusses the hidden scale invariance of Roskilde-simple liquids.  Section \ref{quasicase} restricts the discussion to monatomic liquids and arrives at a new formulation of quasiuniversality. Some consequences of this are presented in Sec. \ref{cons}. Finally, Sec. \ref{out} gives a few concluding remarks.

\section{Background}\label{backgr}

\subsection{Notation}\label{prel}

We consider a classical-mechanical system of $N$ particles of mass $m$ in volume $V$. Sections \ref{gencase} and \ref{hsi} deal with the general case, whereas the discussion of quasiuniversality in Secs. \ref{quasicase} and \ref{cons} is restricted to systems of identical particles. The density $\rho$ is defined by

\be
\rho\equiv\frac{N}{V}\,.
\ee
An example of a Roskilde-simple liquid is the well-known Lennard-Jones (LJ) system defined by the pair potential $v(r)=4\varepsilon[(r/\sigma)^{-12}-(r/\sigma)^{-6}]$, but we do not generally assume pairwise additive forces. Periodic boundary conditions are used throughout. Whenever thermodynamic quantities are referred to, these are always {\it excess} quantities, i.e., having been subtracted the value of the same quantity for an ideal gas at the same density and temperature \cite{han13}.

If the particle positions are denoted by $\br_1,...,\br_N$, the collective $3N$-dimensional position vector $\bR$ is defined by

\be
\bR\equiv (\br_1,...,\br_N)\,.
\ee
An important ingredient in the formulation of quasiuniversality is the use of reduced units. While studies of the LJ liquid traditionally introduce dimensionless lengths and energies by scaling with respect to the pair-potential parameters $\sigma$ and $\varepsilon$, the isomorph theory -- as well as quasiuniversality -- refer to the physics given in {\it macroscopically} reduced units. These are defined as follows: the length unit is $\rho^{-1/3}$, the energy unit is $k_BT$ where $k_B$ is Boltzmann's constant and $T$ the temperature,  and the time unit is $\rho^{-1/3}\sqrt{m/k_BT}$ (for Brownian dynamics a different time unit applies \cite{IV}). As an example, the reduced collective position vector is defined by

\be\label{reddef}
\tbR\equiv \rho^{1/3}\bR\,.
\ee

Any thermodynamic state point is characterized by its constant-potential-energy hypersurface $\Omega$ defined by $\Omega\equiv \{\bR\, |\, U(\bR)=\langle U\rangle\}$ in which $\langle U\rangle$ is the average potential energy at the state point in question. The corresponding reduced-coordinate constant-potential-energy hypersurface is given by

\be\label{Omega_tilde}
\tO\equiv \{\tbR\,|\, U(\rho^{-1/3}\tbR)=\langle U\rangle\}\,.
\ee

\subsection{Isomorphs}\label{iso_sec}

This section reviews the concept of isomorphs introduced in 2009 in Ref. \onlinecite{IV}. An isomorph is a curve in the thermodynamic phase diagram of certain systems -- the Roskilde-simple liquids -- along which several properties in reduced units are invariant to a good approximation.

By scaling of all coordinates a given typical microconfiguration of a thermodynamic state point at one density corresponds uniquely to a microconfiguration at another density; in the case of a molecular system the centers of masses are scaled while each molecule's size and orientation is left unchanged \cite{ing12b}. Two thermodynamic state points are isomorphic if such pairs of scaled microconfigurations have the same canonical probability. In fact, not all pairs must satisfy this, it is enough that all {\it physically relevant} pairs of microconfigurations have same probability {\it to a good approximation}. Here the term ``physically relevant'' does not refer merely to having a reasonable probability -- for instance in a highly viscous liquid a configuration corresponding to a large energy barrier, which is quite unlikely, may well be physically relevant if for instance the barrier is a bottleneck for flow.

Formally, two thermodynamic state points with density and temperature $(\rho_1,T_1)$ and $(\rho_2,T_2)$ are isomorphic if a constant $C_{12}$ exists such that whenever two physically relevant microconfigurations of the state points, $\bR_1\in(\rho_1,T_1)$ and $\bR_2\in(\rho_2,T_2)$, have the same reduced coordinates, i.e., $\rho_1^{1/3}\bR_1=\rho_2^{1/3}\bR_2$, one has

\be\label{iso_def1}
\exp\left(-\frac{U(\bR_1)}{k_BT_1}\right)
\,=\,C_{12}\exp\left(-\frac{U(\bR_2)}{k_BT_2}\right)\,.
\ee
This defines a mathematical equivalence relation in the thermodynamic phase diagram. The equivalence classes are continuous curves, which are the system's isomorphs. The only systems that obey Eq. (\ref{iso_def1}) exactly are those for which the potential energy is a homogeneous function,  i.e., obeys $U(\lambda\bR)=\lambda^{-n}U(\bR)$ for some $n$, for instance systems with inverse-power-law (IPL) pair potentials for which two state points are isomorphic whenever $\rho_1^{n/3}/T_1=\rho_2^{n/3}/T_2$. In this case $C_{12}=1$, but for realistic systems it is found that $C_{12}\neq 1$.

One can rewrite the isomorph definition as follows to express explicitly that the two microconfigurations have the same reduced coordinate $\tbR$:

\be\label{iso_def2}
\exp\left(-\frac{U(\rho_1^{-1/3}\tbR)}{k_BT_1}\right)
\,=\,C_{12}\exp\left(-\frac{U(\rho_2^{-1/3}\tbR)}{k_BT_2}\right)\,.
\ee

As shown in Ref. \onlinecite{IV} the defining identity Eq. (\ref{iso_def1}) implies that several quantities are invariant along an isomorph when given in reduced units. Examples of isomorph invariants are the excess entropy, the isochoric specific heat, the instantaneous shear modulus, the diffusion constant, the viscosity, etc. In fact, the entire dynamics in reduced units is invariant along an isomorph; in particular normalized time-autocorrelation functions, etc, are all invariant. Likewise, structure is invariant in reduced units -- not just the radial distribution function, but all higher-order distribution functions, as well. Examples of quantities that are not isomorph invariant are the reduced-unit free energy, pressure, and  bulk modulus. The fact that many quantities are isomorph invariant means that the thermodynamic phase diagram is effectively one dimensional with respect to these quantities. Of course, since isomorphs are approximate, so are the  isomorph invariants. 

Finally, we mention the concept of an ``isomorph jump'' \cite{IV}: If a system is in thermal equilibrium at one state point, and density and temperature are suddenly changed to those of another, isomorphic state point, the system is instantaneously in thermal equilibrium at the new state point, even if its relaxation time is very large. Thus isomorphs act as a kind of wormholes in the thermodynamic phase diagram.

Which systems have good isomorphs? Recall that the virial $W(\bR)\equiv -1/3 \bR\cdot\nabla U(\bR)$ gives the contribution to pressure from interactions, $pV=Nk_BT+\langle W\rangle$. Appendix A of Ref. \onlinecite{IV} proved the following constitutive theorem characterizing Roskilde-simple liquids, stating that the following three conditions are equivalent:

\begin{enumerate}
\item The system has strong correlations between the equilibrium, constant-volume fluctuations of virial and potential energy.
\item The system has good isomorphs.
\item The system has curves in the thermodynamic phase diagram (the isomorphs) along which the reduced-coordinate constant-potential-energy hypersurfaces $\tO$ are invariant to a good approximation.
\end{enumerate}
Based on extensive computer simulations, Ref. \onlinecite{ing12} suggested a fourth equivalent condition, namely that interactions beyond the first coordination shell may be ignored without changing the physics. The criterion $R>0.9$ for the virial /potential-energy correlation coefficient \cite{ped08} was used in most of our papers as a pragmatic delimitation of the class of Roskilde-simple liquids, but of course the theory does not break down at one particular value of $R$.

Except for systems with a homogeneous potential-energy function, no systems have isomorphs in the whole thermodynamic phase diagram. Nevertheless, extensive computer simulations have shown many realistic systems to be Roskilde simple, for instance \cite{ing12} the Lennard-Jones (LJ) system and its generalizations to mixtures and to other exponents than 6 and 12, simple molecular liquids like the asymmetric dumbbell or the Wahnstrom OTP model, the Buckingham liquid, the ``repulsive'' LJ system (i.e., with plus instead of minus between the two IPL terms). Recently it was shown that the 10-bead flexible LJ chain also has good isomorphs \cite{vel13}, which provides a highly nontrivial example of a Roskilde-simple liquid. The isomorph theory works very well for the crystalline phase; thus an LJ crystal has more than 99\% $WU$ correlations \cite{II,cryst}, which is a nontrivial anharmonic effect that survives in the (classical) zero-temperature limit \cite{II}. In all these cases the theory was checked by tracing out isomorphs in the thermodynamic phase diagram and testing for the predicted invariants. The methods used for generating isomorphs in simulation have been detailed elsewhere \cite{ing12a,boh13}. 

For real liquids, we believe that van der Waals bonded and metallic liquids are generally Roskilde simple, whereas covalently-bonded liquids (e.g., molten silica) and hydrogen-bonded liquids (e.g., alcoholds) are generally not, because strong directional bonding appears to ruin the theory (the flexible LJ chain is a striking exception to this, though). The theory may break down for van der Waals liquids of highly elongated molecules \cite{fra11}. On the other hand, while we previously suggested that ionic liquids are not Roskilde simple, it seems likely that such systems with not too strong Coulomb forces, e.g., room-temperature ionic liquids, may well be so. In experiment, glass-forming Roskilde-simple liquids are characterized by a Prigogine-Defay ratio close to unity \cite{bai08,gun11}; moreover, such liquids obey density scaling and isochronal superposition \cite{IV}, which are characteristic features of van der Waals liquids. 

Note finally that Roskilde-simple liquid have simple thermodynamics. Thus if $s$ is the excess entropy per particle, temperature factorizes as follows \cite{ing12a}

\be\label{thermo}
k_BT=f(s)h(\rho)\,.
\ee 
This equation is mathematically equivalent \cite{ing12a} to the configuration-space version of the well-known Gruneisen high-pressure equation of state expressing that pressure is a linear function of energy with a proportionality constant that depends only on density. A consequence is that, since excess entropy is an isomorph invariant, the isomorphs are given by

\be\label{isomch}
\frac{h(\rho)}{k_BT}={\rm Const.}
\ee

For a Roskilde-simple liquid  the solid-liquid coexistence curve is an isomorph, implying invariance along this curve of a number of quantities like the excess entropy, the reduced viscosity, the reduced radial distribution function, etc \cite{IV,V}.

\subsection{The approximate nature of isomorphs and isomorph invariants}

As mentioned, with the exception of systems with a homogeneous potential-energy function, which do not exist in the real world, isomorphs are only approximate. Any quantity that is defined by reference to thermodynamic state points gives rise to a set of level curves in the two-dimensional thermodynamic phase diagram. If the system is a Roskilde-simple liquid, these curves are almost identical for all the isomorph invariant quantities. Which set of these curves to designate ``isomorphs'' is a matter of taste, but we have consistently {\it defined} the configurational adiabats to be the isomorphs. 

Having emphasized the approximate nature of the isomorph theory, the question is for which systems it works and how well. Here the virial/potential-energy correlation coefficient $R$ is useful. The only case of exact isomorphs, that of a homogenous potential-energy function, is characterized by $R=1$. By continuity one expects that the closer $R$ is to unity, the better are the isomorph invariants. This is confirmed by simulations. Moreover -- also by continuity -- the closer $R$ is to unity, the better an approximation it is to replace the system in question by an inverse power law (IPL) pair potential system; this is also confirmed by simulations \cite{ped10}.

No realistic systems are Roskilde simple in the entire fluid part of the thermodynamic phase diagram. The isomorph theory breaks down when the critical point is approached (where a different kind of simplicity takes over, of course), as well as in the ordinary gas phase. Moreover, the theory quickly breaks down when entering the region of (metastable) states of negative pressures. A typical Roskilde-simple liquid has good isomorphs at all its condensed-phase state points, including the high-temperature, high-pressure supercritical state points not too far from the liquid-solid coexistence line, as well as in the crystalline and the supercooled liquid phases.

\section{Characterizing the potential-energy function}\label{gencase}

This section derives a scaling-type characterization of a Roskilde-simple liquid's potential-energy function. In Sec. \ref{constth} we proceed to derive from this the constitutive theorem and the basic thermodynamic separation identity Eq. (\ref{thermo}).

\subsection{The pseudohomogeneous nature of $U(\bR)$}

This subsection shows that the potential energy function of a Roskilde-simple liquid is characterized by a dimensionless function $\tU(\tbR)$ and two functions with dimension energy, $h(\rho)$ and $g(\rho)$, such that for any physically relevant microconfiguration $\bR$ of a state point with density $\rho$ one has

\be\label{prop1m}
U(\bR)\cong h(\rho)\tU(\tbR)+g(\rho)\,.
\ee
As becomes clear in Sec. \ref{constth}, the function $h(\rho)$ is the same as that appearing in the basic thermodynamic separation identity Eq. (\ref{thermo}). 

The reduced-coordinate constant-potential-energy hypersurfaces $\tO$ defined in Eq. (\ref{Omega_tilde}) are generally high-dimensional Riemannian manifolds parameterized by the two thermodynamic coordinates. For a Roskilde-simple liquid, however, these hypersurfaces are given by just one parameter (Sec. \ref{iso_sec}) \cite{IV}. This is indicated by writing (where the parameter $\lambda$ in one-to-one correspondence with $h(\rho)/(k_BT)$, compare Eq. (\ref{isomch}))

\be
\tO(\lambda)\,.
\ee
The excess quantities $S$ and $C_V$ are both determined by the manifold $\tO(\lambda)$ \cite{dyr13}: In the microcanonical ensemble the excess entropy $S$ is the logarithm of the area of $\tO(\lambda)$; the proof that the excess isochoric heat capacity $C_V$ is also encoded in $\tO(\lambda)$ is given in Ref. \onlinecite{dyr13}. Combining the facts that $S=S(\lambda)$ and $C_V=C_V(\lambda)$ with the identity $C_V=(\partial S/\partial\ln T)_\rho$, one has for variations at constant density $\rho$ if $\phi(\lambda)\equiv S'(\lambda)/C_V(\lambda)$ 

\be\label{dlnt1}
d\ln T = \phi(\lambda)d\lambda\,.
\ee
Integrating this leads to $\ln T(\lambda,\rho)-\ln T(\lambda_0,\rho)=\int_{\lambda_0}^\lambda \phi(\lambda)d\lambda$. Thus the ratio $T(\lambda,\rho)/ T(\lambda_0,\rho)$ is independent of $\rho$, which means that for any $\rho_1$ and $\rho_2$ one has $T(\lambda,\rho_1)/T(\lambda_0,\rho_1)=T(\lambda,\rho_2)/T(\lambda_0,\rho_2)$. Consequently, if $H(\rho_1,\rho_2)\equiv T(\lambda_0,\rho_1)/T(\lambda_0,\rho_2)$ one has for all $\lambda$

\be\label{T121}
T(\lambda,\rho_1)=H(\rho_1,\rho_2)\, T(\lambda,\rho_2)\,.
\ee

Now suppose that $\tbR\in\tO(\lambda)$. If $\tbR_0$ is an arbitrary point on a reference manifold $\tO(\lambda_0)$, the potential-energy change from manifold $\tO(\lambda_0)$ to $\tO(\lambda)$ at constant density $\rho_1$ is given by $\int_{\lambda_0}^{\lambda}C_V(\lambda)dT(\lambda)$ in which $C_V(\lambda)$ depends only on the manifold $\tO(\lambda)$ and not on density. The potential-energy change is therefore given by

\be\label{deltaU1}
U(\rho_1^{-1/3}\tbR) - U(\rho_1^{-1/3}\tbR_0) =\int_{\lambda_0}^{\lambda}C_V(\lambda)\frac{\partial T(\lambda,\rho_1)}{\partial\lambda}d\lambda\,.
\ee
Although Eq. (\ref{deltaU1}) refers to single microconfigurations, it calculates the potential-energy difference by a thermodynamic integration argument. This unusual procedure makes sense because all microconfigurations on a constant-potential-energy hypersurface of course have the same potential energy. -- From Eqs. (\ref{T121}) and (\ref{deltaU1}) we conclude that 

\be\label{eq14}
U(\rho_1^{-1/3}\tbR) - U(\rho_1^{-1/3}\tbR_0)
= H(\rho_1,\rho_2) \left[ U(\rho_2^{-1/3}\tbR) - U(\rho_2^{-1/3}\tbR_0)\right]\,.
\ee
Note that from this one easily recovers Eq. (\ref{T121}) via the well-known configurational temperature expression \cite{LLstat,rug97,pow05} 

\be\label{configtemp}
k_BT = \frac{\langle (\nabla U)^2\rangle}{\langle \nabla^2 U\rangle}\,.
\ee

Since the reference quantities $U(\rho_1^{-1/3}\tbR_0)$ and $U(\rho_2^{-1/3}\tbR_0)$ are functions of $\rho_1$ and $\rho_2$, we conclude from Eq. (\ref{eq14}) that one can write for a suitable function $G(\rho_1,\rho_2)$

\be \label{prop1}
U(\rho_1^{-1/3}\tbR)=H(\rho_1,\rho_2)U(\rho_2^{-1/3}\tbR)+G(\rho_1,\rho_2)\,.
\ee
If $\rho_0$ is a reference density, Eq. (\ref{T121}) implies $T(\lambda_0,\rho_1)/T(\lambda_0,\rho_0)=H(\rho_1,\rho_0)$ and $T(\lambda_0,\rho_2)/T(\lambda_0,\rho_0)=H(\rho_2,\rho_0)$. Defining $h(\rho)\equiv H(\rho,\rho_0)$ this implies via Eq. (\ref{T121}) that $H(\rho_1,\rho_2)=h(\rho_1)/h(\rho_2)$ (using a different reference density corresponds to multiplying $h(\rho)$ by a constant). Combining this with Eq. (\ref{prop1}) we conclude that whenever $\rho_1^{1/3}\bR_1=\rho_2^{1/3}\bR_2(\equiv\tbR)$, one has

\be\label{Uh1}
U(\bR_1)
\cong {h(\rho_1)}\frac{U(\bR_2)}{h(\rho_2)}+G(\rho_1,\rho_2)\,.
\ee
In order to emphasize the approximate nature of the theory we have replaced the equality sign by a $\cong$ sign. 

Define 

\be\label{tudef1}
\tU(\tbR)\equiv\frac{U(\rho_0^{-1/3}\tbR)}{h(\rho_0)}\,.
\ee
Because the function $h(\rho)$ is only defined within an overall multiplicative constant, the same applies for $\tU(\tbR)$. 
Substituting into Eqs. (\ref{Uh1}) $\rho_1=\rho$ and $\rho_2=\rho_0$ and combining with Eq. (\ref{tudef1}) we finally arrive at Eq. (\ref{prop1m}) in which $g(\rho)\equiv G(\rho,\rho_0)$.

Two equivalent ways of writing Eq. (\ref{prop1m}) are

\be\label{prop3m}
U(\bR)\cong h(\rho)\tU(\rho^{1/3}\bR)+g(\rho)
\ee
and

\be\label{prop4m}
U(\rho^{-1/3}\tbR)\cong h(\rho)\tU(\tbR)+g(\rho)\,.
\ee
For a homogeneous potential-energy function, for instance a system of IPL pair potentials $\propto r^{-n}$, one has $U(\rho^{-1/3}\tbR)=\rho^{n/3}\tU(\tbR)$, i.e., $h(\rho)\propto\rho^{n/3}$ and $g(\rho)=0$. Equation (\ref{prop4m}) is a generalization of this, and one may thus say that Roskilde-simple liquids are characterized by a ``pseudohomogeneous'' potential-energy function that obeys Eq. (\ref{prop4m}).

\begin{figure}[H]
  \centering
  \includegraphics[width=80mm]{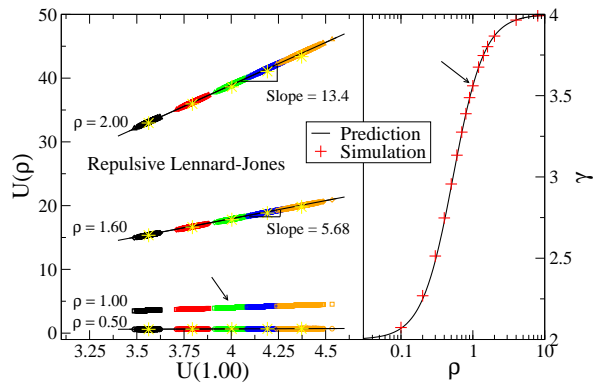}
  \caption{Results from a simulation of $1000$ particles of the repulsive LJ fluid defined by the pair potential $(r^{-12}+r^{-6})/2$. The left panel shows the potential energies of pairs of instantaneous microconfigurations, where the potential energy of a given microconfiguration at the simulated density $1.0$ is denoted $U(1.00)$ and that of the same microconfiguration scaled to density $\rho$ is denoted $U(\rho)$ ($\rho=0.5; 1.6; 2.0$). This was done for $T=0.6; 0.8; 1.0; 1.2; 1.4$ (black, red, green, blue, yellow). The black lines have slopes determined from the fluctuations calculated at the state point $(\rho,T)=(1,1)$ marked by an arrow (see the text). The right panel shows the density-scaling exponent $\gamma$ predicted from $h(\rho)$ for state points along one isomorph (full curve) compared to $\gamma$ calculated at each state point (red crosses). The arrow again marks the state point $(\rho,T)=(1,1)$. Reprinted with permission from T. S. Ingebrigtsen, L. B{\o}hling, T. B. Schr{\o}der, and J. C. Dyre, J. Chem. Phys. {\bf 136}, 061102 (2012). Copyright 2011, AIP Publishing LLC.}
  \label{fig2}
\end{figure}

As an illustration we show in Fig. 1 results \cite{ing12a} from simulations of a system of $1,000$ particles interacting via the ``repulsive'' LJ pair potential defined $v(r)=(r^{-12}+r^{-6})/2$. The repulsive LJ fluid extrapolates between an $n=6$ IPL behavior at very low densities to an $n=12$ IPL behavior at very high densities; this system has $R>0.99$ apparently everywhere in the phase diagram. The left panel  shows instantaneous values of the potential energy scaled to different densities for each of five temperatures marked by separate colors, plotted versus the instantaneous potential energy at $\rho=1$ where the simulation was carried out. The fact that these scatter plots are highly elongated demonstrates that the proportionality of Eq. (\ref{Uh1}) is a good approximation. The lines were calculated from the expression $h(\rho)=A\rho^2+B\rho^4$ (see Refs. \onlinecite{ing12a} or \onlinecite{boh12}, an expression that is also derived below from quasiuniversality), in which only the ratio $A/B$ is relevant. This number was determined from simulations at the state point $(\rho,T)=(1,1)$ marked by the arrow. The right panel shows the variation along the isomorph through $(\rho,T)=(1,1)$ of the so-called density-scaling exponent $\gamma\equiv (\partial\ln T/\partial\ln\rho)_S$ \cite{IV}, which from Eq. (\ref{thermo}) is given by 

\be\label{gamlign}
\gamma = \frac{d\ln h}{d \ln\rho}\,.
\ee
The full curve gives $\gamma$ calculated from $h(\rho)=A\rho^2+B\rho^4$ \cite{ing12a} and the red crosses mark $\gamma$ calculated from the fluctuation expression $\gamma=\langle\Delta W\Delta U\rangle/\langle(\Delta U)^2\rangle$ \cite{IV}.

\subsection{Deriving the constitutive theorem from the pseudohomogeneous property of $U(\bR)$}\label{constth}

Equation (\ref{prop1m}) boils down the physics of a Roskilde-simple liquid into one equation. It is an approximate identity that does not apply for all $\bR$, but for the physically relevant microconfigurations of state points corresponding to typical condensed-liquid states, i.e., not too far from the solid-liquid coexistence line, as well as all crystalline states \cite{cryst}. This section shows how the fundamentals of the isomorph theory -- the three constitutive properties characterizing a Roskilde-simple liquid, the thermodynamic separation idenity Eq. (\ref{thermo}) (Sec. \ref{iso_sec}), and Eq. (\ref{gamlign}) -- all follow in a straightforward manner from Eq. (\ref{prop1m}).

1. The virial $W(\bR)\equiv (-1/3) \bR \cdot {\bf \nabla} U(\bR)$ determines how much the potential energy changes per relative density change if a microconfiguration is scaled uniformly \cite{tildesley}, i.e., if its reduced coordinates are kept constant. It is easy to show that

\be
W(\bR)= \left(\frac{\partial U(\bR)}{\partial\ln\rho}\right)_{\tbR}\,.
\ee
From Eq. (\ref{prop1m}) we thus get $W(\bR)\cong (d h / d\ln\rho)\tU(\tbR)+ d g/ d\ln\rho$. Eliminating $\tU(\tbR)$ in this expression via Eq. (\ref{prop1m}) leads to

\be\label{WUrel1}
W(\bR)\cong\gamma(\rho)U(\bR)+\phi(\rho)\,,
\ee
in which $\gamma(\rho)\equiv d\ln h/d\ln\rho$ (Eq. (\ref{gamlign})) and $\phi(\rho)=-\gamma(\rho)g(\rho)+dg/d\ln\rho$. Equation (\ref{WUrel1}) implies that at any fixed density virial and potential energy are strongly correlated. In particular, for constant-density equilibrium fluctuations at one state point one has $\Delta W(\bR)\cong\gamma\, \Delta U(\bR)$ \cite{I,II} in which $\gamma$ depends only on the density \cite{IV}.

2. Equation (\ref{prop1m}) further implies the existence of isomorphs. Consider two densities, $\rho_1$ and $\rho_2$. For any positive number $K$ one can define two temperatures $T_1$ and $T_2$ by $k_BT_1=K\,h(\rho_1)$ and $k_BT_2=K\,h(\rho_2)$. Then Eq. (\ref{Uh1}), which trivially follows from Eq. (\ref{prop1m}), implies Eq. (\ref{iso_def1}), i.e., the two state points $(\rho_1,T_1)$ and $(\rho_2,T_2)$ are isomorphic. The constant $K$ identifies the isomorph, which implies that an isomorph is the set of state points given by Eq. (\ref{isomch}). In fact, since isomorphic state points have the same excess entropy, $K$ is a function of this, $K=f(s)$, which straight away gives the thermodynamic separation identity Eq. (\ref{thermo}). -- We note that it is possible to derive a differential-geometric expression for the function $f(s)$. Substituting Eqs. (\ref{prop3m}) into Eq. (\ref{configtemp}) leads to $k_BT = h(\rho){\langle ( \tilde\nabla \tU)^2\rangle}/{\langle \tilde\nabla^2 \tU\rangle}$, which implies

\be\label{configtemp2}
f(s) = \frac{\langle (\tilde\nabla \tU)^2\rangle}{\langle \tilde\nabla^2 \tU\rangle}\,. 
\ee
In this expression the ensemble averages may be calculated as canonical averages or as configuration-space microcanonical averages, i.e., by integrating over the relevant manifold $\tO$.

3. To derive the third part of the constitutive theorem, we note that in terms of the function $\tilde\Phi(\tbR)$ via Eq. (\ref{prop1m}) the manifolds $\tO$ are given by 

\be\label{tourel}
\tO=\{\tbR\,|\,\tU(\tbR)={\rm Const.}\}\,.
\ee
Consequently, this family is parameterized by a single number.

\section{Hidden scale invariance}\label{hsi}

The underlying cause of isomorph invariance is what we have termed {\it hidden scale invariance} \cite{sch09,III}. Consider for simplicity a liquid of particles interacting via the pair potential $v(r)$. By dimensional analysis an energy $\varepsilon$ and a length $\sigma$ exist such that one can write $v(r)=\varepsilon\, \phi(r/\sigma)$ in which $\phi$ is a dimensionless function. In the case of an IPL pair potential one has $\phi(x)=x^{-n}$. Scale invariance is the mathematical condition $\phi(\lambda x)=\lambda^{-n}\phi(x)$. Physically, scale invariance implies that there is no characteristic energy or length of the potential; the only relevant quantity is their combination $\varepsilon\sigma^n$ that is the prefactor of $r^{-n}$. We shall now see how these two properties generalize to Roskilde-simple liquids' potential-energy function.

For any system, any reduced quantity $\tilde A$ is dimensionless and therefore a function of the two dimensionless variables $\trh\equiv\rho\sigma^3$ measuring the density and $\tT\equiv k_BT/\varepsilon$ measuring the temperature: $\tilde A=\tilde A(\trh,\tT)$. For a Roskilde-simple liquid a function $f(\trh,\tT)$ exists that determines $\tilde A$ in the following way: $\tilde A(\trh,\tT)=\tilde A (f(\trh,\tT))$, in which the function $f(\trh,\tT)$ is common to all the isomorph invariants of the given system. This function's level curves in the phase diagram are the isomorphs, of course. 

The above means that the reduced-unit physics of a Roskilde-simple liquid does not depend separately on energy scale or spatial range of the potential, but only on a {\it combination} of these two quantities. In this sense the system has a hidden scale invariance, reminiscent of the well-known scale invariance of critical phenomena. Hidden scale invariance is  only approximate, of course, but so is the standard scale invariance close to the critical point. The difference is that the latter involves a power-law scaling that becomes more and more accurate as the critical point is approached, whereas hidden scale invariance involves a more general form of scaling that nowhere becomes exact.

For any system of classical particles one can write for some dimensionless function $\tV$

\be\label{cls}
U(\bR)=\varepsilon\,\tV(\bR/\sigma)\,,
\ee
where $\varepsilon$ sets the energy scale of the interactions and $\sigma$ their length scale. The hidden scale invariance of a Roskilde-simple liquid manifests itself in Eq. (\ref{prop3m}). In this approximate identity, by dimensional analysis (see, e.g., Ref. \onlinecite{jhj13} and its references) the function $h(\rho)$, which has dimension energy, may for some dimensionless function $\tilde{h}$ be written $h(\rho)=\varepsilon \tilde{h}(\rho\sigma^3)$ and likewise for $g(\rho)$. Thus

\be\label{prop5m}
U(\bR)\cong \varepsilon \left[\tilde{h}(\rho\sigma^3)\tU(\rho^{1/3}\bR)+\tilde{g}(\rho\sigma^3)\right]\,.
\ee
This provides a precise definition of hidden scale invariance. Structure and dynamics in reduced units are determined entirely by the function $\tU(\rho^{1/3}\bR)$ that has no intrinsic length scale. At the same time, the overall energy scale is not set by $\varepsilon$, but by the {\it combination} of the microscopic energy and length scales, $\varepsilon\,\tilde{h}(\rho\sigma^3)$, just like the prefactor in the IPL case. In particular, the physics at the state point $(\rho,T)$ is determined by the single number $\varepsilon {h}(\rho\sigma^3)/k_BT=\tilde{h}(\tilde\rho)/\tT$, which leads to Eq. (\ref{isomch}) characterizing an isomorph. For small density variations the function $h(\rho)$ can be approximated well by a power law, $h(\rho)\cong\rho^\gamma$, bringing hidden scale invariance close to conventional scale invariance.

\section{Quasiuniversality}\label{quasicase}

This section gives a potential-energy formulation of quasiuniversality, supplementing the previous formulation in terms of the manifolds $\tO(\lambda)$ being quasiuniversal \cite{dyr13}. We consider only single-component liquids of atomic particles, i.e., with no intermolecular structure. The most important case is that of particles interacting via pairwise additive forces, but this assumption is not necessary.

\subsection{Quasiuniversality of simple systems}\label{Quasi}

We first briefly summarize the quasiuniversalities reported in the literature during the last 50 years, which were to a large extent justified via computer simulations. The reader is referred to Ref. \onlinecite{dyr13} for more details and further references. The quasiuniversalities include

\begin{itemize}

\item Different IPL systems' close similarities with respect to structure and dynamics, similarities that extend to all other Roskilde-simple liquids \cite{ash66,han69,hoo71,sti75,ros76,gue03,bra06,hey09,lan09,ram09,sch11,pon11a,ram11,lop12}.

\item The Young-Andersen approximate scaling principle, stating that if two liquids at two state points have the same reduced-unit radial distribution function, they have the same reduced-unit dynamics \cite{you03,you05}.

\item Quasiuniversality of the translational/orientational order-parameter maps of Debenedetti and coworkers \cite{tru00,err03,cha07}.

\item Excess entropy scaling, stating that the reduced-unit diffusion constants of different liquids have an approximately universal dependence on the excess entropy per particle \cite{ros77,ros99,pon11,sin12}.

\item The Lindemann melting criterion \cite{ubb65} and seven other quasiuniversal melting/freezing rules \cite{dyr13}.

\item Quasiuniversality of Roskilde-simple liquids' specific-heat temperature dependence \cite{ing13}, all conforming to the Rosenfeld-Tarazona expression \cite{ros99} $C_V \propto T^{-2/5}$.

\item Quasiuniversal isochoric fragility \cite{alb02,dem04,nis07}.

\end{itemize}

All quasiuniversalities have exceptions, but it appears that these always involve systems that are not Roskilde-simple. The following properties of monatomic Roskilde-simple liquids were derived from the $\tO$ formulation of quasiuniversality of Ref. \onlinecite{dyr13}, i.e., that the manifolds $\tO(\lambda)$ are quasiuniversal:

\begin{itemize}

\item The basic thermodynamic separation property Eq. (\ref{thermo}) is partly quasiuniversal,

\be\label{EOS}
k_B T=f_0(s)h(\rho)\,,
\ee 
in which the function $f_0(s)$ is common to all Roskilde-simple liquids whereas $h(\rho)$ is not.

\item Additivity: If $U_1(\bR)$ and $U_2(\bR)$ are potentials of two monatomic Roskilde-simple liquids, so is $U_1(\bR)\pm U_2(\bR)$ whenever this is a well-defined system, i.e., with a potential-energy function that has a lower bound.

\item Quasiuniversal interdependence of any two isomorph invariants, generalizing excess entropy scaling.

\item A single microconfiguration is enough to determine all isomorph invariants of the thermodynamic state point in question without knowing the Hamiltonian.

\end{itemize}

\subsection{Quasiuniversality and hidden scale invariance}

Some of the above quasiuniversalities manifest themselves in the fact that a single number determines the physics. This is the case for instance for excess entropy scaling or the vibrational mean-square displacement controlling melting in the Lindemann criterion; likewise the collapse of order-parameter maps for different liquids to a quasiuniversal curve implies that the translational order parameter determines the rotational. And the standard liquid-state perturbation theories for the radial distribution function regard a liquid as a hard-sphere system to zeroth order \cite{zho09} -- here the hard-sphere packing fraction is the single parameter controlling the physics.

Suppose that quasiuniversality applies and is reflected in the fact that some parameter, $X$, determines the physics. Then the curves of constant $X$ in the thermodynamic phase diagram have invariant physics. These curves generally involve both density and temperature variations. This implies hidden scale invariance for quasiuniversal liquids in the sense that neither the energy nor the length scales of Eq. (\ref{cls}) separately determine the physics, which is instead given by a combination of those quantities. In other words, quasiuniversality implies hidden scale invariance. This explains why quasiuniversality appears to be limited to Roskilde-simple liquids. It also {\it suggests} -- albeit in no way {\it implies}, of course -- the opposite, that all monatomic Roskilde-simple liquids are quasiuniversal.

\subsection{Potential-energy formulation of quasiuniversality}

Reference \onlinecite{dyr13} showed that all the above listed quasiuniversal properties can be derived from the fact/postulate that all monatomic Roskilde-simple liquids have the same reduced-coordinate constant-potential-energy hypersurfaces $\tO(\lambda)$. Combining this with Eq. (\ref{prop1m}) suggests that these liquids have in common the functions $\tU(\tbR)$, because according to Eq. (\ref{prop3m}) this would immediately imply that they have they have the family $\tO(\lambda)$ in common. 

A possible argument against this is the following. If for instance systems $1$ and $2$ obey $\tU_1(\tbR)=\exp(\tU_2(\tbR))$, they have the same system of constant-potential-energy hypersurfaces, but not proportional potential-energy functions. This would never happen in reality, though, because it would violate the fact that interactions are local in $3d$ space. Assuming that the potential energy is a sum of independent contributions from different parts of space, by a generalization of the proof leading to Eq. (\ref{prop1m}) it is possible to show \cite{dyr13a} that, indeed, the function $\tU(\tbR)$ is common to all monatomic Roskilde-simple liquids. This means that one can write

\be\label{kvas}
U(\bR)\cong h(\rho){\tilde\Phi}_0(\tbR)+g(\rho)\,,
\ee
in which the function ${\tilde\Phi}_0(\tbR)$ is quasiuniversal whereas $h(\rho)$ and $g(\rho)$ are not. 

Equation (\ref{kvas}) is not to be interpreted in a strict sense. Thus one aspect of the potential energy function is not quasiuniversal, namely the form of the interaction between two particles forced into close contact. For the case of a pair-potential liquid the probability of some small separation $r$ between two particles is roughly $\exp(-v(r)/k_BT)$, which clearly violates quasiuniversality. This affects the steepness of the pair distribution function $g(r)$ below its first maximum and the height of its first maximum. We have found in simulations that the collective physical properties are virtually unaffected, however, and these are the ones in focus here. 

The correct interpretation of Eq. (\ref{kvas}) is that all quasiuniversalities are easily derived from this equation. Moreover, all of the physics that is {\it not} quasiuniversal, e.g., the equation of state, the reduced pressure, the reduced free energy, the reduced bulk modulus, etc, are consequences of the fact that the functions $h(\rho)$ and $g(\rho)$ are system specific. The criterion for which quantities are quasiuniversal and which are not, is that the former are precisely the isomorph invariant quantities.

\section{Some new consequences of Eqs. (\ref{prop3m}) and (\ref{kvas})}\label{cons}

As mentioned, it is straightforward to show that all of the quasiuniversalities listed in Sec. \ref{Quasi} follow from Eq. (\ref{kvas}). Regarding novel consequences, note that the isomorph concept, as well as that of strong virial / potential-energy correlations, both relate to properties of equilibrium thermodynamic state points. Roskilde-simple liquids' quasiuniversality, however, applies beyond thermal equilibrium. Below we give brief examples of this, but first we consider the analytical structures of the nonuniversal functions $h(\rho)$ and $g(\rho)$.

\subsection{The analytical structures of $h(\rho)$ and $g(\rho)$}\label{hogg}

It was shown in Refs. \onlinecite{ing12a} and \onlinecite{boh12} that for a Roskilde-simple liquid with pair potential $v(r)=\sum_n \varepsilon_n (r/\sigma)^{-n}$, the function $h(\rho)$ is given by an expression of the form $h(\rho)=\sum_n C_n \rho^{n/3}$ in which the only non-zero terms are those that correspond to terms in $v(r)$. This follows from Eq. (\ref{thermo}) \cite{ing12a}, but as shown below it also follows from quasiuniversality, which further implies the same analytical structure for $g(\rho)$. 

Consider first an IPL system, $v_n(r)=\varepsilon_n (r/\sigma)^{-n}$; its corresponding functions are denoted $h_n(\rho)$ and $g_n(\rho)$. For fixed $\tbR$ the potential energy scales with density as $U(\bR)\propto \rho^{n/3}$. Comparing this to Eq. (\ref{prop3m}) we conclude that $h_n(\rho){\tilde\Phi}_0(\tbR)+g_n(\rho)\propto\rho^{n/3}$ for several values of ${\tilde\Phi}_0(\tbR)$. This implies $h_n(\rho)\propto\rho^{n/3}$ and $g_n(\rho)\propto\rho^{n/3}$. For dimensional reasons we can thus write $h_n(\rho)=\alpha_n \varepsilon_n(\rho \sigma^3)^{n/3}$ and $g_n(\rho)=\beta_n \varepsilon_n(\rho \sigma^3)^{n/3}$ in which $\alpha_n$ and $\beta_n$ are numerical constants. 

In the general case, $v(r)=\sum_n \varepsilon_n (r/\sigma)^{-n}$, the potential energy is a sum of IPL terms, $U(\bR)=\sum_n U_n(\bR)$. For each IPL term we have according to Eq. (\ref{kvas}) $U_n(\bR)\cong h_n(\rho){\tilde\Phi}_0(\tbR)+g_n(\rho)$. Adding these we get $U(\bR)\cong h(\rho){\tilde\Phi}_0(\tbR)+g(\rho)$ in which $h(\rho)=\sum_n \alpha_n \varepsilon_n(\rho \sigma^3)^{n/3}$ and $g(\rho)=\sum_n \beta_n \varepsilon_n(\rho \sigma^3)^{n/3}$. Thus the functions $h(\rho)$ and $g(\rho)$ both inherit the analytical structure of $v(r)$.

The free-energy variation along an isomorph is determined by the constant $C_{12}$ of Eq. (\ref{iso_def1}). It follows from Eq. (\ref{prop3m}) that $C_{12}$ is determined by $g(\rho)$. Thus quasiuniversality provides a prediction for the free energy's variation along an isomorph. It would be interesting to check this by simulation, e.g., for the LJ system for which $g(\rho)=C \rho^4-D\rho^2$ for some $C$ and $D$.

\subsection{Extending the isomorph concept}\label{ext}

Isomorphs are defined by reference to the canonical ensemble and to the physically relevant microconfigurations of thermodynamic equilibrium states \cite{IV}. Equation (\ref{prop3m}) not just applies for microconfigurations that are typical at some definite temperatures, however. For instance, Eq. (\ref{prop3m}) can be used to rationalize computer simulations of zero-temperature plastic flows of LJ-type liquids because the function $h(\rho)$ sets the energy scale of the flow events \cite{ler13}. Another application of Eq. (\ref{prop1m}) is for extending the isomorph concept to nonlinear flows described by the SLLOD equation of motion \cite{sep13}. In this case the relevant phase diagram acquires a third dimension defined by the shear rate. The density-temperature relation of the SLLOD-generalized isomorphs turned out not to involve the shear rate \cite{sep13}, however, a non-trivial fact that follows from Eq. (\ref{prop3m}).

\subsection{Quantum liquids}\label{ql}

Consider a Roskilde-simple liquid with potential energy function $U(\bR)$. The Schr{\"o}dinger equation referring to density $\rho_1$ is written

\be\label{schr1}
\left(-\frac{\hbar^2}{2m}\nabla_{\bR_1}^2+U(\bR_1)\right)\left.\left|\Psi^{(1)}_n\right.\right>
=E^{(1)}_n \left.\left|\Psi^{(1)}_n\right.\right>\,.
\ee
Likewise at density $\rho_2$ we write

\be\label{schr2}
\left(-\frac{\hbar^2}{2m}\nabla_{\bR_2}^2+U(\bR_2)\right)\left.\left|\Psi^{(2)}_n\right.\right>=E^{(2)}_n\left.\left|\Psi^{(2)}_n\right.\right>\,.
\ee
Multiplying Eq. (\ref{schr1}) by $h(\rho_2)/h(\rho_1)$ and using Eq. (\ref{prop1m}) leads to

\be\label{schr3}
\left(-\frac{\hbar^2}{2m}\frac{h(\rho_2)}{h(\rho_1)}\,\nabla_{\bR_1}^2+U(\bR_2)+h(\rho_2)g(\rho_1,\rho_2)\right)\left.\left|\Psi^{(1)}_n\right.\right>\cong E^{(1)}_n\frac{h(\rho_2)}{h(\rho_1)} \,\left.\left|\Psi^{(1)}_n\right.\right>\,.
\ee
Comparing Eqs. (\ref{schr2}) and (\ref{schr3}) and using $\rho_1^{1/3}\bR_1=\rho_2^{1/3}\bR_2$, which implies 
$\nabla_{\bR_1}^2=(\rho_2/\rho_1)^{-2/3}\nabla_{\bR_2}^2$, shows how to relate the spectrum at density $\rho_2$ to that at density $\rho_1$ (where the eigenvalue dependence of mass and density are given explicitly and 
$m'\equiv m\, h(\rho_1) \rho_1^{-2/3} /[h(\rho_2) \rho_2^{-2/3}]$:

\be\label{sch1}
E_n\left(m', \rho_2\right)
\cong\frac{h(\rho_2)}{h(\rho_1)}E_n(m,\rho_1)-h(\rho_2)g(\rho_1,\rho_2)\,.
\ee
The corresponding (unnormalized) eigenfunctions obey (where $\rho_1^{1/3}\bR_1=\rho_2^{1/3}\bR_2$)

\be
\Psi^{(1)}_n(\bR_1)\cong\Psi^{(2)}_n(\bR_2)\,.
\ee
Equation (\ref{sch1}) can be written in a symmetric form as follows

\be\label{sch2}
\frac{E_n\left(m_1, \rho_1\right)-E_0\left(m_1, \rho_1\right)}{h(\rho_1)}
\,\cong\,\frac{E_n\left(m_2, \rho_2\right)-E_0\left(m_2, \rho_2\right)}{h(\rho_2)}\,
\ee
in which

\be
\frac{m_1}{m_2}
\,=\,\frac{h(\rho_1)\rho_1^{-2/3}}{h(\rho_2)\rho_2^{-2/3}}\,.
\ee
Thus changing density gives a new spectrum, which is the old spectrum displaced and scaled for a system with a different mass. In other words, if the solution to the Schr{\"o}dinger equation is known for one density, it is known for all densities.

Equation (\ref{sch2}) applies for any Roskilde-simple liquid, and due to quasiuniversality the ratio appearing in Eq. (\ref{sch2}) will be the same for all Roskilde-simple liquids.

\section{Concluding remarks}\label{out}

This paper established a precise formulation of the hidden scale invariance of Roskilde-simple liquids (Eq. (\ref{prop1m})) and showed how this leads to a simple potential-energy formulation of the quasiuniversality of single-component atomic such liquids (Eq. (\ref{kvas})).

According to Eq. (\ref{prop1m}) the reduced-unit structure and dynamics are controlled by the dimensionless term $\tU(\tbR)$, which does not involve any microscopic length or energy. It is not trivial that many realistic systems to a good approximation have this form of generalized or ``hidden'' scale invariance, in which the intrinsic length scale $\sigma$ of the potential plays no role. Note that this does no mean that $\sigma$ is physically insignificant -- for instance the $\sigma$ of the LJ pair potential gives the typical interparticle distance at low and moderate pressure.

We argued briefly above that quasiuniversality implies the hidden scale invariance that lies behind the isomorph concept. Quasiuniversality is (presumably) restricted to the class of monatomic Roskilde-simple liquids, and the isomorph concept remains central because it generalizes to multicomponent and molecular systems \cite{ing12b}, for instance the apparently  quite complex system like the flexible LJ chain \cite{vel13}.

What is the connection between different monatomic Roskilde-simple liquids' isomorphs? Each system's isomorphs are labeled by the (excess) entropy. The actual shape of the isomorphs in the thermodynamic phase diagram is not quasiuniversal, it depends on the nonuniversal functions $h(\rho)$ and $g(\rho)$. But quasiuniversality implies that a unique mapping exists between different systems' isomorphs such that for isomorphs with same entropy all other isomorph invariants are also identical.

\acknowledgments 

The author is indebted to Nick Bailey for critical reading of the manuscript. The centre for viscous liquid dynamics ``Glass and Time'' is sponsored by the Danish National Research Foundation via grant DNRF61.

\end{document}